# Ultrafast mid-infrared nanoscopy
# of strained vanadium dioxide nanobeams

M. A. Huber[1], M. Plankl[1], M. Eisele[1], R. E. Marvel[2], F. Sandner[1], T. Korn[1], C. Schüller[1], R. F. Haglund, Jr.[2], R. Huber[1], T. L. Cocker[1]

[1] *Department of Physics, University of Regensburg, 93040 Regensburg, Germany*
[2] *Department of Physics and Astronomy and Interdisciplinary Materials Science Program, Vanderbilt University, Nashville, Tennessee 37235-1807, USA*

# Ultrafast mid-infrared nanoscopy of strained vanadium dioxide nanobeams

M. A. Huber[1], M. Plankl[1], M. Eisele[1], R. E. Marvel[2], F. Sandner[1], T. Korn[1], C. Schüller[1], R. F. Haglund, Jr.[2], R. Huber[1], T. L. Cocker[1]

[1] *Department of Physics, University of Regensburg, 93040 Regensburg, Germany*
[2] *Department of Physics and Astronomy and Interdisciplinary Materials Science Program, Vanderbilt University, Nashville, Tennessee 37235-1807, USA*

**Long regarded as a model system for studying insulator-to-metal phase transitions, the correlated electron material vanadium dioxide ($VO_2$) is now finding novel uses in device applications. Two of its most appealing aspects are its accessible transition temperature (~341 K) and its rich phase diagram. Strain can be used to selectively stabilize different $VO_2$ insulating phases by tuning the competition between electron and lattice degrees of freedom. It can even break the mesoscopic spatial symmetry of the transition, leading to a quasi-periodic ordering of insulating and metallic nanodomains. Nanostructuring of strained $VO_2$ could potentially yield unique components for future devices. However, the most spectacular property of $VO_2$—its ultrafast transition—has not yet been studied on the length scale of its phase heterogeneity. Here, we use ultrafast near-field microscopy in the mid-infrared to study individual, strained $VO_2$ nanobeams on the 10 nm scale. We reveal a previously unseen correlation between the local steady-state switching susceptibility and the local ultrafast response to below-threshold photoexcitation. These results suggest that it may be possible to tailor the local photo-response of $VO_2$ using strain and thereby realize new types of ultrafast nano-optical devices.**

The insulator-to-metal phase transition in vanadium dioxide (VO$_2$) has been the subject of extensive investigation since its discovery in 1959 (Ref. 1). Interest has stemmed in part from its relatively simple, nonmagnetic structure[2] and its accessible transition temperature (T$_c$ ~ 341 K), which makes it relevant for technological applications[3-6]. Nevertheless, the enduring appeal of VO$_2$ can be traced to the complex interplay between electron and lattice degrees of freedom that produce its intricate free energy landscape[7-12]. This fine balance between competing interactions can be tuned by strain, leading to a rich phase diagram[13,14]. Below T$_c$, unstrained VO$_2$ is an insulator characterized by both strong electron-electron correlations and lattice distortion, where the vanadium ions form chains of dimerized pairs (monoclinic structure, M1). These dimers are dissociated in the rutile (R), metallic phase (T > T$_c$) in a process reminiscent of a Peierls transition. However, density functional theory calculations have shown that both lattice distortion and strong on-site electron-electron Coulomb repulsion are necessary to accurately model the band gap[9]. Moreover, tensile strain applied to the insulating state can produce new, stable lattice structures that are intermediates between M1 and R, and, surprisingly, these states are also insulators. High tensile strain along the rutile $c_R$ axis (>2%, along the direction of the dimerized chains in the insulating state) induces the monoclinic insulating phase M2 in which only every second row of vanadium ions is dimerized[15,16]. Meanwhile, moderate tensile strain results in the triclinic T phase, which exhibits a phonon spectrum that is tunable with strain, yet still distinct from both M1 and M2[15,16]. Structurally, the T phase is a continuous deformation of the M1 phase. It persists with increasing tensile strain until the material undergoes a discontinuous transition to the M2 phase[16].

Ultrafast studies exploring the complementary, photoinduced phase transition in VO$_2$ have sought to separate its competing electron and lattice degrees of freedom in the time domain[17-32]. When photoexcited by an ultrafast near-infrared pump pulse, insulating VO$_2$ can be driven into the metallic state via a non-thermal route[21,26-29] that is faster than the electron-lattice thermalization time. Studies of this process have highlighted a threshold carrier population for switching[11,27] and the impact of coherent structural oscillations between the monoclinic and rutile lattices[20,21,26-28,30]. Terahertz (far-infrared) and multi-terahertz (mid-infrared) spectroscopy[13,21,22,24,26,27,33-38] have played key roles in unraveling the ultrafast dynamics of $VO_2$. These wavelengths probe the complex conductivity at energies less than the bandgap of the insulating state ($E_g$~0.6-0.7 eV), where a large, persistent free-carrier response is direct evidence for the metallic phase. Ultrafast far- and mid-infrared spectroscopy can even reveal the photoinduced formation of the metallic phase, or alternatively the perturbative response of the insulating phase to below-threshold photoexcitation[21,26,27]. Here, ultrafast near-infrared photodoping creates a transient population of carriers in the conduction and valence bands of the insulating state that survives for less than 1 ps (Refs. 21,26). This semiconducting response manifests itself as a quickly decaying free carrier conductivity of moderate strength in the terahertz and multi-terahertz regimes[21,26,27]. Terahertz spectroscopy has also been used to study metallic domain evolution through the conductivity signatures of carrier localization[22,27,36,38]. This has yielded insights into the average dynamics of nanodomains, but the local influence of defects, crystallinity and strain have remained out of reach because far-field spectroscopy is an inherently macroscopic technique. Meanwhile, scattering-type near-field scanning optical microscopy (s-NSOM) in the far- and mid-infrared has proven uniquely suited to investigating the local behavior of the steady-state transition[32,35,39-44]. Since s-NSOM at these frequencies accesses the nanoscale plasmonic response[40] (and thus the local free-carrier conductivity) strong scattering contrast is observed between the insulating and metallic states. As a result, near-field images of $VO_2$ thin films and crystals show clearly the phase coexistence and domain growth typical of a first-order phase transition. The inherent sensitivity of s-NSOM in the far- and mid-infrared makes it the ideal tool to characterize new types of $VO_2$ samples, from strained films[42,43] to single micro- and nanocrystals[32,41], though ultrafast s-NSOM[45-47] has not yet been applied to $VO_2$.

Single micro- and nanocrystals offer the least complex, and thus most fundamental, model materials in which to access the VO$_2$ phase transition. Single-crystal VO$_2$ studies were rare until recently because large single crystals typically fracture upon thermal cycling. Nevertheless, the growth of robust single microcrystals by physical vapor transport (PVT) is now well established[48,49]. Interestingly, in the case of single-crystal nanobeams, PVT also introduces significant, non-epitaxial strain between VO$_2$ and the substrate. This strain arises due to the high growth temperature and the different thermal expansion coefficients of VO$_2$ and the substrate. VO$_2$ nanobeams are generally grown at temperatures on the order of 1000°C, and hence naturally form in the metallic state, where the rutile $c_R$ axis is oriented along the long axis of the nanobeam. Since the thermal expansion coefficient of metallic VO$_2$ along the $c_R$ axis is usually larger than that of the substrate[41,50,51], tensile strain is introduced upon cooling towards the transition temperature. An added complication arises as the sample passes below the transition temperature because the lattice constants of the various phases of VO$_2$ are all different along the $c_R$ direction ($\ell_R < \ell_{M1} < \ell_T < \ell_{M2}$). Therefore, depending on the precise growth conditions and substrate, an insulating nanobeam can be stabilized in the M1, T, or M2 phase[41,50], or even in the M1 phase under compressive strain[51]. Substrate-induced strain leads to a striking effect when the nanobeam is subsequently heated back above the transition temperature: quasi-periodic metallic nanodomains form along the nanobeam axis. This phase-coexistence structure persists to temperatures above the normal phase transition region. In the literature this effect has been described as a dynamic equilibrium, balancing the energies of strain, domain wall formation, and entropy (phase switching)[41,50-52]. Here, we use ultrafast near-field mid-infrared microscopy to show that while the concept of strain balancing is essentially sound, the domain periodicity is actually built into the insulating state for a nanobeam stabilized in the T phase by substrate clamping. Specifically, the spatial variation of the strain in the insulating state is conferred to the local, below-threshold photoconductivity. This implies that the domain structure during phase coexistence is not purely the result of a dynamic energy balance in the heated direction, but rather is predetermined, likely by a freezing-in of the periodicity in the initial cooling stage after nanobeam growth.

The VO$_2$ nanobeams investigated in this work were grown by PVT on a *c*-cut sapphire substrate. Vanadium pentoxide powder was heated to 810°C in a quartz crucible and the resulting gas was flowed over the substrate, which was also maintained at 810°C, for 30 minutes. Nanobeam growth proceeded such that the rutile $c_R$ axis for a given nanobeam was parallel to its long axis[50-52]. Notably, the surface of the *c*-cut sapphire substrate introduces an additional, six-fold symmetry, where the nanobeams preferentially grow along one of three directions that are separated by 120°, as can be seen in Figure 1a and 1b (also see Supplementary Figure 1). Additionally, many nanobeams contain a kink and a change to the long axis direction. In a given arm of a kinked nanobeam, the $c_R$ axis remains oriented along the long axis of the arm, so the kink must contain a change to the lattice structure, such as a dislocation.

The nanobeams in the sample investigated here are strained due to the mismatch between thermal expansion coefficients, as was discussed previously. The particular strain state of a given nanobeam can be determined by identifying its insulating phase using Raman spectroscopy. Figure 1c shows a typical Raman spectrum of the highest-energy phonon mode (denoted $\omega_O$) in VO$_2$, which is associated with V-O vibration[15,16,41], recorded in the center of a nanobeam. The phonon mode exhibits the distinctive double-peak shape of the T phase, where the high-frequency peak position is consistent with a nanobeam under ~0.6% tensile strain[16]. In Figure 1d, the frequency of this peak is mapped as a function of space with ~1 μm resolution. Details of the scanning Raman setup are published elsewhere[53]. We find that the strain state of the nanobeam in the center of Figure 1d is not completely uniform. Instead, the yellow areas indicate regions where the tensile strain is slightly lower, leading to a peak shift of ~1 cm$^{-1}$. This corresponds to a reduction in tensile strain by approximately 0.03% (Ref. 16). In contrast, the nanobeam at the top of Figure 1d is almost unstrained (M1 phase), indicating that it has been decoupled from the substrate[41,50,51]. The topography of the central nanobeam, measured by atomic force microscopy, is shown in Figure 2a. The nanobeam is approximately 500 nm wide and 200 nm high, and features a roughly 3-μm-long flat section of single-crystalline VO$_2$ (indicated by white lines in Figure 2a). Interestingly, the ends of this single crystal correspond to the regions of relaxed strain in Figure 1d. Aside from the single crystal, the nanobeam also contains substructure in the form of three grain boundaries, while small crystallites of VO$_2$ decorate the nanobeam in some places.

We investigated the phase transition and ultrafast response of the nanobeam locally using mid-infrared s-NSOM. Our setup was described previously in Ref. 46. Briefly, near-infrared (1560 nm) femtosecond pulses from an ultrastable Er:fibre laser system are tailored spectrally and temporally and used to generate the pump and probe pulses for the ultrafast measurements. Phase-stable multi-THz probe pulses are produced by difference frequency generation in a GaSe crystal (center frequency: 31.5 THz; bandwidth: 10 THz FWHM; duration: 60 fs FWHM), while near-infrared pump pulses (center wavelength: 780 nm; duration: 150 fs FWHM) are obtained by frequency doubling the fundamental laser pulses in a periodically poled lithium niobate crystal. Both the pump and probe pulses are focused onto the tip of a commercial s-NSOM and the intensity of the probe pulses scattered from the s-NSOM tip apex is measured with a mercury cadmium telluride (MCT) photodiode. The s-NSOM is operated in tapping mode, enabling background-free detection of the scattered near-field intensity at the harmonics of the tip tapping frequency (see, e.g., Ref. 54). Using this technique, mid-infrared spatial resolution down to ~20 nm has been shown at the second harmonic of the tapping frequency[40,54] (scattered intensity $I_2$), while 10 nm resolution has been demonstrated at the third harmonic[46] (scattered intensity $I_3$). Additionally, we can study steady-state processes on the 10-nm scale using a mid-infrared quantum cascade laser as a source of continuous probe radiation ($\lambda$ = 8.35 μm, $f$ = 36 THz).

The steady-state scattered near-field intensity from the nanobeam under continuous-wave probe illumination is shown in Figure 2b. At room temperature (top) the response from the single crystal is uniform, while the grain boundaries show up as regions of reduced scattering efficiency. Conversely, the heated nanobeam (bottom) contains areas of high scattering efficiency (metallic phase), most notably periodic domains in the single crystalline horizontal section of the nanobeam (period ~770 nm). In previous studies, it has been speculated that similar domains arise from a strain-induced shift of the transition temperature and a dynamic energy balance, in which each new metallic domain introduces additional tensile strain because the metallic lattice constant along the $c_R$ axis is shorter than that of any of the insulating phases[41,50-52]. The persistence of insulating domains to relatively high temperatures, both in Figure 2b and in previous studies, is consistent with this theory since tensile strain is known to increase the transition temperature. On the other hand, if the domain periodicity is purely a consequence of the increase in substrate-induced strain with each new metallic domain, then the insulating phase should contain no hint of this periodicity. Yet, the Raman spectroscopy map in Figure 1d shows that some substructure is in fact present in the insulating state of our nanobeam, though nanoscale spatial resolution is necessary to explore the insulating nanobeam on the length scale of domain periodicity.

Ultrafast mid-infrared nanoscopy provides a unique perspective into the local character of the insulating nanobeam. An ultrafast near-infrared pump pulse with a focal spot diameter that is larger than the nanobeam excites it uniformly (pump polarization: perpendicular to the shaft of the s-NSOM tip). The ultrafast local response to photoexcitation is subsequently probed on the nanoscale. Multi-THz probe pulses scattered from the s-NSOM tip apex reveal nanobeam substructure via the local evolution of the infrared conductivity. Free carriers dominate the $VO_2$ conductivity in the frequency range of our probe pulse[43], so the near-field scattering dynamics provide a measure of the transient photoexcited carrier population. More specifically, the scattered mid-infrared intensity increases when the local plasma frequency exceeds the center frequency of the probe pulse[46] (31.5 THz). Figure 3a shows the scattered multi-THz intensity as a function of pump-probe delay time, where the s-NSOM tip was positioned near the end of the single crystal. A ~150 fs increase and subpicosecond decay are observed for pump fluences below the phase transition threshold (black points, curve). These dynamics are a signature of a semiconducting response: The ultrafast near-infrared pump pulse populates the conduction and valence bands of the insulating state, increasing the local plasma frequency and hence the scattered multi-THz intensity. The carrier population then decays by electron-hole recombination, which leads to a decay of the scattered near-field intensity. Notably, the time scale of this decay is extremely fast compared to conventional semiconductors because recombination is aided by carrier self-trapping[21,55]. The data in Figure 3a constitute the first ultrafast measurement of the electronic conductivity of $VO_2$ on the nanoscale.

Conversely, for pump fluences above the phase transition threshold (Figure 3a red points, curve), the scattered multi-THz intensity increases drastically at all delay times (Figure 3b). This new equilibrium results from a discontinuous increase in free carrier conductivity and is a fingerprint of the insulator-to-metal phase transition. No ultrafast dynamics are observed for high fluences because the time between pump pulses (50 ns) is shorter than the recovery time of the photoinduced metallic state. Figure 3c shows the emergence of metallic domains in the nanobeam with increasing pump fluence, while Figure 3d shows the corresponding near-field line scans through the center of the nanobeam. The domain positions are reproducible and consistent with the quasi-periodic structure observed at high temperature (Figure 2b). In other words, the threshold pump fluence for phase switching depends on location. The threshold is lower and the switching susceptibility is higher in places where a metallic domain forms. A spatially dependent switching susceptibility in a nanobeam under global excitation could stem from dynamic strain balancing, as has been proposed previously[41,50-52], or strain substructure built into the insulating nanobeam, as strain applied to the insulating state has been shown to increase the $VO_2$ phase transition temperature.

In Figure 4a, regions of high switching susceptibility are marked on the insulating nanobeam, which is visualized via the room-temperature steady-state near-field image (top image in Figure 2b). To explore potential substructure in the insulating beam, particularly in the single crystal, we measured the local transient response to ultrafast below-threshold photoexcitation at room temperature. Near-infrared pump / near-field multi-THz probe traces were recorded for each susceptibility domain defined in Figure 4a. The perturbative conductivity dynamics for the first three domains (P1 – P3) are given in Figure 4b. In all regions the photoinduced carrier density decays almost completely within 1 ps. The distinction between domains is most prominent in the *magnitude* of the photoinduced conductivity dynamics. The peak change to the scattered near-field multi-THz intensity is significantly higher in domains with higher switching susceptibility. We explore this relationship in more detail in a line scan along the nanowire axis (Figure 4c). Remarkably, the magnitude of the perturbative, semiconducting response qualitatively reproduces the periodic domain structure observed when the insulating and metallic states coexist (Figure 2b, Figure 3c,d). We infer a new piece of information from this finding: the periodic metallic domain structure that emerges at high temperatures is built into the insulating state. This conclusion contradicts the hypothesis that metallic domains form upon heating solely due to a dynamic energy balance. We therefore propose a modified model for domain formation, though one based on the same physical principles. Since the nanobeam is grown at 810°C, it is initialized in the fully metallic phase. As it subsequently cools to room temperature the nanobeam passes through the phase coexistence state and insulating nanodomains form spontaneously, possibly aided by nucleation sites[32]. Later, as the nanobeam nears room temperature, the remaining metallic domains vanish and the entire beam is stabilized in the insulating state. Crucially, the insulating domains that form at high temperature (in the phase coexistence regime) do so under different strain conditions than the sections that undergo the phase change near room temperature. The domain periodicity that forms in the initial cooling stage is therefore frozen into the insulating state via the local strain profile. Upon re-heating, metallic domains naturally form first in regions that are preconditioned to support them, as the local strain determines the local switching susceptibility.

Periodic strain in the insulating state provides a potential link between the local steady-state switching susceptibility and the ultrafast semiconducting response. It has been proposed that tensile strain along the $c_R$ axis increases electronic correlations in the insulating state, thereby blue-shifting the bandgap and moving the electronic system further from the conductivity edge[10,13]. We therefore expect that regions with larger semiconducting responses are less strained than their counterparts. This is consistent with our observation that these regions coincide with areas of high steady-state switching susceptibility, since tensile strain has been shown to increase the phase transition temperature of $VO_2$ (Ref. 41). It is also consistent with the Raman spectroscopy map of the nanobeam (Figure 1d). Although the Raman data does not have sufficient spatial resolution to show the complete strain substructure of the nanobeam, regions of reduced strain are visible at the ends of the single crystal, where both the steady-state switching susceptibility and the below-threshold photoconductivity are large.

Finally, we address the possible role of nucleation sites with a complementary experiment (see Supplementary Figure 2). We investigated a polycrystalline $VO_2$ nanoparticle that displayed a clear metallic-domain nucleation site in steady-state near-field imaging of the phase transition. This nanoparticle has dimensions comparable to those of a single domain in the nanobeam, and we therefore expect that strain-induced substructure should not be present in its insulating state. Using ultrafast mid-infrared microscopy, we found that the magnitude of its semiconducting response is spatially uniform within the noise levels of our measurement, and note in particular that the nucleation site did not exhibit an enhanced perturbative photoconductivity. These observations support the idea that strain is essential in connecting switching susceptibility to the ultrafast perturbative response.

In conclusion, we have used ultrafast mid-infrared nanoscopy to reveal heterogeneous local dynamics in a $VO_2$ nanobeam. We have found that the transient local semiconducting response is linked to the steady-state switching susceptibility. This connection most likely results from a nanoscale strain profile that is frozen into the insulating state during cooling following PVT. The capacity to tailor the ultrafast response of $VO_2$ on the nanoscale using strain could lead to numerous future device applications, perhaps the most spectacular being the prospect of clearing and resetting its ultrafast local response with heat. In a hypothetical future experiment, it would be interesting to "anneal" the sample by increasing the nanobeam temperature until it becomes fully metallic and then apply a defined strain profile via the substrate to deterministically set the domain positions. Ultrafast mid-infrared nanoscopy will prove invaluable in characterizing the operation of such nano-optical devices. More fundamentally, the application of field-resolved ultrafast nanoscopy with sub-cycle time resolution[46] to $VO_2$ could provide access to coherent oscillations on the nanoscale, allowing this important aspect of the phase transition to be studied under the influence of confinement and strain.

**Acknowledgements**

The authors thank M. Furthmeier for technical assistance and J. M. Atkin and M. K. Liu for discussions. This work was supported by the European Research Council through ERC grant 305003 (QUANTUMsubCYCLE); the Deutsche Forschungsgemeinschaft through Research Training Group GRK 1570, SFB 689, and Research Grants CO1492/1 and HU1598/3; and the National Science Foundation (DMR-1207507). T.L.C. acknowledges the support of the A. v. Humboldt Foundation.

**References**

(1) Morin, F. J. *et al. Phys. Rev. Lett.* **1959**, 3 , 34-36.

(2) Goodenough, J. B. *J. Solid State Chem*. **1971**, 3, 490-500.

(3) Nakano, M.; Shibuya, K.; Okuyama, D.; Hatano, T.; Ono, S.; Kawasaki, M.; Iwasa, Y.; Tokura, Y. *Nature* **2012**, 487, 459-462.

(4) Jeong, J.; Aetukuri, N.; Graf, T.; Schladt, T. D.; Samant, M. G.; Parkin, S. S. P. *Science* **2013**, 339, 1402-1405.

(5) Aetukuri, N. B.; Gray, A. X.; Drouard, M.; Cossale, M.; Gao, L.; Reid, A. H.; Kukreja, R.; Ohldag, H.; Jenkins, C. A.; Arenholz, E.; Roche, K. P.; Dürr, H. A.; Samant, M. G.; Parkin, S. S. P. *Nature Phys*. **2013**, 9, 661-666.

(6) Ryckman, J. D.; Hallman, K. A.; Marvel, R. E.; Haglund, R. F., Jr.; Weiss, S. M. *Opt. Express* **2013**, 21, 10753-10763

(7) Zylbersztejn, A.; Mott, N. F. *Phys. Rev. B* **1975**, 11, 4383-4395.

(8) Wentzcovitch, R. M.; Schulz, W. W; Allen, P. B. *Phys. Rev. Lett.* **1994**, 72 3389-3392.

(9) Biermann, S.; Poteryaev, A.; Lichtenstein, A. I.; Georges, A. *Phys. Rev. Lett*. **2005**, 94, 026404.

(10) Lazarovits, B.; Kim, K.; Haule, K.; Kotilar, G. *Phys. Rev. B* **2010**, 81, 115117.

(11) van Veenedaal, M. *Phys. Rev. B* **2013**, 87, 235118.

(12) Budai, J. D.; Hong, J.; Manley, M. E.; Specht, E. D.; Li, C. W.; Tischler, J. Z.; Albernathy, D. L.; Said, A. H.; Leu, B. M.; Boatner, L. A.; McQueeney, R. J.; Delaire, O. *Nature* **2014**, 515, 535-542.

(13) Abreu, E.; Liu, M.; Lu, J.; West, K. G.; Kittiwatanakul, S.; Yin, W.; Wolf, S. A.; Averitt, R. D. *New J. Phys.* **2012**, 14, 083026.

(14) Park, J.-H.; Coy, J. M.; Kasirga, T. S.; Huang, C.; Fei, Z.; Hunter, S.; Cobden, D. H. *Nature* **2013**, 500, 431-434.

(15) Marini, C.; Arcangeletti, E.; Di Castro, D.; Baldassare, L.; Perucchi, A.; Lupi, S.; Malavasi, L.; Boeri, L.; Pomjakushina, E.; Conder, K.; Postorino, P. *Phys. Rev. B* **2008**, 77, 235111.

(16) Atkin, J. A.; Berweger, S.; Chavez, E. K.; Raschke, M. B.; Cao, J.; Fan, W.; Wu, J. *Phys. Rev. B* **2012**, 85, 020101(R).

(17) Basov, D. N.; Averitt, R. D.; van der Marel, D.; Dressel, M.; Haule, K. *Rev. Mod. Phys*. **2011**, 83, 471-541.

(18) Becker, M. F.; Bruckman, A. B.; Walser, R. M.; Lépine, T.; Georges, P.; Brun, A. *Appl. Phys. Lett*. **1994**, 65, 1507-1509.

(19) Cavalleri, A.; Dekorsy, Th.; Chong, H. H. W.; Kieffer, J. C.; Schoenlein, R. W. *Phys. Rev. B*. **2004**, 70, 161102(R).

(20) Kim, H.-T.; Lee, Y. W.; Kim, B.-J.; Chae, B.-G.; Yun, S. J.; Kang, K.-Y.; Han, K.-J.; Yee, K.-J.; Lim, Y.-S. *Phys. Rev. Lett*. **2006**, 97, 266401.

(21) Kübler, C.; Ehrke, H.; Huber, R.; Lopez, R.; Halabica, A.; Haglund, R. F., Jr.; Leitenstorfer, A. *Phys. Rev. Lett.* **2007**, 99, 116401.

(22) Hilton, D. J.; Prasankumar, R. P.; Fourmaux, S.; Cavalleri, A.; Brassard, D.; El Khakani, M. A.; Kieffer, J. C.; Taylor, A. J.; Averitt, R. D. *Phys. Rev. Lett.* **2007**, 99, 226401.

(23) Baum, P.; Yang, D.-S.; Zewail, A. H. *Science* **2007**, 318, 788-792.

(24) Nakajima, M.; Takubo, N.; Hiroi, Z.; Ueda, Y.; Suemoto, T. *Appl. Phys. Lett.* **2008**, 92, 011907.

(25) Hada, M.; Okimura, K.; Matsuo, J. *Phys. Rev. B* **2010**, 82, 153401.

(26) Pashkin, A.; Kübler, C.; Ehrke, H.; Lopez, R.; Halabica, A.; Haglund, R. F., Jr.; Huber, R.; Leitenstorfer, A. *Phys. Rev. B* **2011**, 83, 195120.

(27) Cocker, T. L.; Titova, L. V.; Fourmaux, S.; Holloway, G.; Bandulet, H.-C.; Brassard, D.; Kieffer, J.-C.; El Khakani, M. A.; Hegmann, F. A. *Phys. Rev. B* **2012**, 85, 155120.

(28) Wall, S.; Wegkamp, D.; Foglia, L.; Appavoo, K.; Nag, J.; Haglund, R. F., Jr., Stähler, J.; Wolf, M. *Nat. Commun.* **2012**, 3, 721.

(29) Wegkamp, D.; Herzog, M.; Xian, L.; Gatti, M.; Cudazzo, P.; McGahan, C. L.; Marvel, R. E.; Haglund, R. F., Jr; Rubio, A.; Wolf, M.; Stähler, J. *Phys. Rev. Lett.* **2014**, 113, 216401.

(30) Appavoo, K.; Wang, B.; Brady, N. F.; Seo, M.; Nag, J.; Prasankumar, R. P.; Hilton, D. J.; Pantelides, S. T.; Haglund, R. F., Jr. *Nano Lett.* **2014**, 14, 1127-1133.

(31) Morrison, V. R.; Chatelain, R. P.; Tiwari, K. L.; Hendaoui, A.; Bruhács, A.; Chaker, M.; Siwick, J. *Science* **2014**, 346, 445-448.

(32) O'Callahan, B. T.; Jones, A. C.; Park, J. H.; Cobden, D. H.; Atkin, J. M.; Raschke, M. B. *Nature Commun.* **2015**, 6, 6849.

(33) Tonouchi, M. *Nature Photon.* **2007**, 1, 97-105.

(34) Jepsen, P. U.; Fischer, B. M.; Thoman, A.; Helm, H.; Suh, J. Y.; Lopez, R.; Haglund, R. F., Jr. *Phys. Rev. B* **2006**, 74, 205103.

(35) Zhan, H.; Astley, V.; Hvasta, M.; Deibel, J. A.; Mittleman, D. M.; Lim, Y.-S.; *Appl. Phys. Lett.* **2007**, 91, 162110.

(36) Cocker, T. L.; Titova, L. V.; Fourmaux, S.; Bandulet, H.-C., Brassard, D.; Kieffer, J.-C.; El Khakani, M. A.; Hegmann, F. A. *Appl. Phys. Lett.* **2010**, 97, 221905.

(37) Liu, M.; Hwang, H. Y.; Tao, H.; Strikwerda, A. C.; Fan, K.; Keiser, G. R.; Sternbach, A. J.; West, K. G.; Kittiwatanakul, S.; Lu, J.; Wolf, S. A.; Omenetto, F. G.; Zhang, X.; Nelson, K. A.; Averitt, R. D. *Nature* **2012**, 487, 345-348.

(38) Lourembam, J.; Srivastava, A.; La-o-vorakiat, C.; Rotella, H.; Venkatesan, T.; Chia, E. E. M. *Sci. Rep.* **2015**, 5, 9182.

(39) Atkin, J. M.; Berweger, S.; Jones, A. C.; Raschke, M. B. *Adv. Phys.* **2012**, 61, 745-842.

(40) Qazilbash, M. M.; Brehm, M.; Chae, B.-G.; Ho, P.-C.; Andreev, G. O.; Kim, B.-J.; Yun, S. J.; Balatsky, A. V.; Maple, M. B.; Keilmann, F.; Kim, H.-T.; Basov, D. N. *Science* **2007**, 318, 1750-1753.

(41) Jones, A. C.; Berweger, S.; Wei, J.; Cobden, D.; Raschke, M. B. *Nano Lett.* **2010**, 10, 1574-1581.

(42) Liu, M. K.; Wagner, M.; Abreu, E.; Kittiwatanakul, S.; McLeod, A.; Fei, Z.; Goldflam, M.; Dai, S.; Fogler, M. M.; Lu, J.; Wolf, S. A.; Averitt, R. D.; Basov, D. N. *Phys. Rev. Lett.* **2013**, 111, 096602.

(43) Liu, M.; Sternbach, A. J.; Wagner, M.; Slusar, T. V.; Kong, T.; Bud'ko, S. L.; Kittiwatanakul, S.; Qazilbash, M. M.; McLeod, A.; Fei, Z.; Abreu, E.; Zhang, J.; Goldflam, M.; Dai, S.; Ni, G.-X.; Lu, J.; Bechtel, H. A.; Martin, M. C.; Raschke, M. B.; Averitt, R. D.; Wolf, S. A.; Kim, H.-T.; Canfield, P. C.; Basov, D. N. *Phys. Rev. B* **2015**, 91, 245155.

(44) Abate, Y.; Marvel, R. E.; Ziegler, J. I.; Gamage, S.; Javani, M. H.; Stockman, M. I.; Haglund, R. F., Jr. *Sci. Rep.* **2015**, 5, 13997.

(45) Wagner, M.; Fei, Z.; McLeod, A. S.; Rodin, A. S.; Bao, W.; Iwinski, E. G.; Zhao, Z.; Goldflam, M.; Liu, M.; Dominguez, G.; Thiemens, M.; Fogler, M. M.; Castro Neto, A. H.; Lau, C. N.; Amarie, S.; Keilmann, F.; Basov, D. N. *Nano Lett.* **2014**, 14, 894-900.

(46) Eisele, M.; Cocker, T. L.; Huber, M. A.; Plankl, M.; Viti, L.; Ercolani, D.; Sorba, L.; Vitiello, M. S.; Huber, R. *Nature Photon.* **2014**, 8, 841-845.

(47) Wagner, M.; McLeod, A. S.; Maddox, S. J.; Fei, Z.; Liu, M.; Averitt, R. D.; Fogler, M. M.; Bank, S. R.; Keilmann, F.; Basov, D. N. *Nano Lett.* **2014**, 14, 4529-4534.

(48) Strelcov, E.; Davydov, A. V.; Lanke, U.; Watts, C.; Kolmakov, A. *ACS Nano* **2011**, 5, 3373-3384.

(49) Kim, I. S.; Lauhon, L. J. *Cryst. Growth Des.* **2012**, 12, 1383-1387.

(50) Wu, J.; Gu, Q.; Guiton, B. S.; de Leon, N. P.; Ouyang, L.; Park, H. *Nano Lett.* **2006**, 6, 2313-2317.

(51) Wei, J.; Wang, Z.; Chen, W.; Cobden, D. H. *Nature Nanotech.* **2009**, 4, 420-424.

(52) Cao, J.; Ertekin, E.; Srinivasan, V.; Fan, W.; Huang, S.; Zheng, H.; Yim, J. W. L.; Khanal, D. R.; Ogletree, D. F.; Grossman, J. C.; Wu, J. *Nature Nanotech.* **2009**, 4, 732-737.

(53) Plechinger, G.; Mann, J.; Preciado, E.; Barroso, D.; Nguyen, A.; Eroms, J.; Schüller, C.; Bartels, L.; Korn, T. *Semicond. Sci. Technol.* **2014**, 29, 064008.

(54) Stiegler, J. M.; Huber, A. J.; Diedenhofen, S. L.; Gómez Rivas, J.; Algra, R. E.; Bakkers, E. P. A. M.; Hillenbrand, R. *Nano Lett.* **2010**, 10, 1387-1392.

(55) Dexheimer, S. L.; Van Pelt, A. D.; Brozik, J. A.; Swanson, B. I. *Phys. Rev. Lett.* **2000**, 84, 4425.

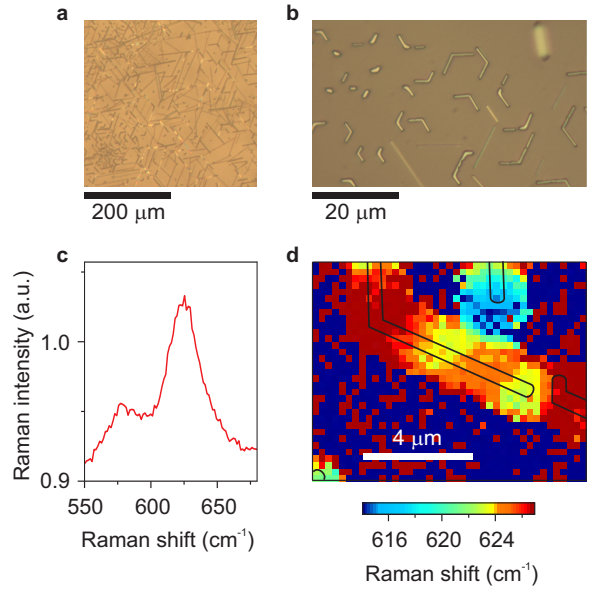

FIGURE 1. (a), (b) Optical microscope images of vanadium dioxide (VO$_2$) nanobeams grown on a *c*-cut sapphire substrate. The nanobeams are strongly bound to the substrate and the $c_R$ axis of the VO$_2$ rutile phase is oriented along the long axis of a given nanobeam. (c) Raman spectrum from a VO$_2$ nanobeam in the spectral range of the $\omega_O$ phonon mode of VO$_2$ showing the distinctive double-peak response characteristic of the triclinic (T) insulating phase. (d) False-color map of the high-frequency peak position of the triclinic $\omega_O$ phonon mode revealed by spatially resolved Raman spectroscopy (spatial resolution ~1 μm). Raman spectra were recorded at room temperature with a probe fluence of ~1 mW. Black lines: sketch of actual nanobeam profiles.

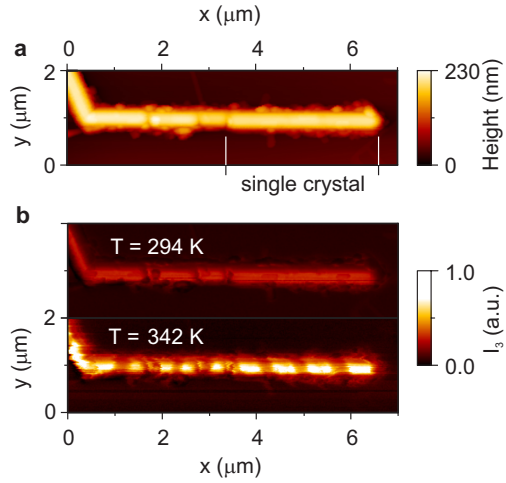

FIGURE 2. (a) Atomic force microscope (AFM) image of the nanobeam in the center of Figure 1(d). The white lines indicate an extended region of single crystalline $VO_2$. (b) Scattering-type near-field scanning optical microscopy (s-NSOM) images of the nanobeam recorded using a continuous wave probe at a wavelength of 8.35 μm. Top: nanobeam at a temperature of T = 294 K, below the phase transition temperature. Bottom: nanobeam at T = 342 K (±1 K), above the transition temperature. Bright, periodic domains form upon heating, indicating a local phase transition from the insulating to the metallic state.

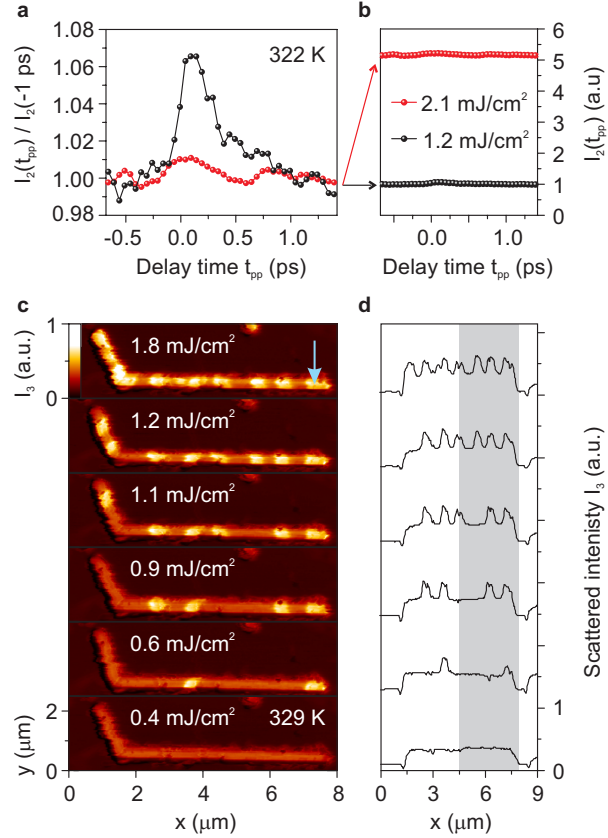

FIGURE 3. Response of the VO$_2$ nanobeam to photoexcitation by an ultrafast near-infrared pump pulse. (a) The multi-terahertz (multi-THz) intensity scattered from the end of the beam at 322 K exhibits a sharp peak and a subpicosecond decay as a function of pump-probe delay (pump: 780 nm, 150 fs; probe: 31.5 THz, 60 fs) provided the pump fluence is below the phase transition threshold (black points, curve). For higher pump fluences the transient response is suppressed (red points, curve), while the average scattering efficiency is increased by a factor of five (b), indicating that a steady-state metallic phase has been formed. (c) The transition threshold depends on location, and domain formation along the nanobeam can be observed in steady-state s-NSOM images measured as a function of pump fluence (base temperature: 329 K). Blue arrow: s-NSOM tip location for (a) and (b). (d) Averaged line cuts through the center of the nanobeam illustrate the periodic nature of the domain structure, which displays a repetition over approximately 770 nm in the single crystal region (grey shaded area).

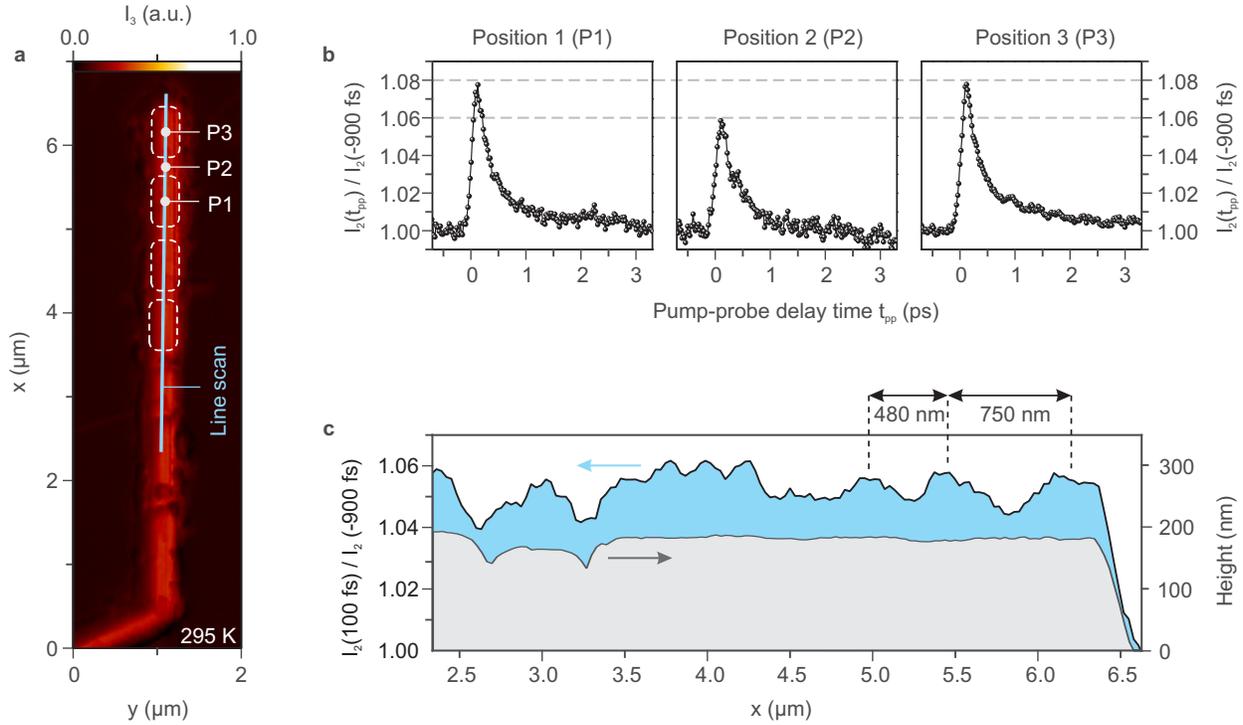

FIGURE 4. Spatially resolved ultrafast response below the phase transition threshold. At room temperature the steady-state near-field mid-infrared scattering response (a) along the axis of the nanobeam is approximately uniform in the single-crystal region, in contrast to its periodic structure at 342 K (illustrated by white dashed lines outlining the metallic domain locations in Figure 2(b)). However, the local response to ultrafast near-infrared photoexcitation is not uniform at 295 K. (b) Near-infrared pump / near-field multi-THz probe traces near the end of the nanobeam (at Positions P1, P2 and P3) exhibit distinct differences depending on the local steady-state switching susceptibility (pump fluence: 2.1 mJ/cm²). The peak of the photoinduced scattering response is larger in regions where the phase transition can readily occur (P1, P3). (c) A line scan along the nanobeam axis at $t_{pp}$ = 100 fs pump-probe delay shows the fine spatial structure of the peak height of the transient response. It qualitatively follows the periodic metallic domain structure observed for the nanobeam heated (Figure 2(b)) and optically pumped (Figure 3(c) and (d)) above the transition. On the other hand, a topographic line scan taken simultaneously is approximately flat in the single-crystal region.